\documentstyle[emulateapj,epsf]{article}

\begin{document}

\def\deg{^{\circ}}

\newcommand{\psrae}{\mbox{J0737$-$3039A}}
\newcommand{\psra}{\mbox{J0737$-$3039A }}
\newcommand{\psrbe}{\mbox{J0737$-$3039B}}
\newcommand{\psrb}{\mbox{J0737$-$3039B }}
\newcommand{\be}{\begin{eqnarray}}
\newcommand{\ee}{\end{eqnarray}}

\title{The Double Pulsar System J0737$-$3039: \\ Modulation 
of A by B at eclipse} \author{M.\ A.\ McLaughlin\altaffilmark{1}, A. G.
Lyne\altaffilmark{1}, D.\ R.\ Lorimer\altaffilmark{1}, A.
Possenti\altaffilmark{2}, \\ R.\ N.\
Manchester\altaffilmark{3},  F. Camilo\altaffilmark{4}, I. H. Stairs\altaffilmark{5},
M.\ Kramer\altaffilmark{1}, \\ 
 M.\ Burgay\altaffilmark{2},  
N.\ D'Amico\altaffilmark{6}, P. C. C.
Freire\altaffilmark{7},  B. C. Joshi\altaffilmark{8} \& N.\ D.\ R.\ Bhat\altaffilmark{9}}

\altaffiltext{1}{Jodrell Bank Observatory, University of Manchester, Macclesfield, Cheshire, SK11 9DL, UK}
\altaffiltext{2}{INAF - Osservatorio Astronomico di Cagliari, Loc. Poggio dei
Pini, Strada 54, 09012 Capoterra, Italy}
\altaffiltext{3}{Australia Telescope National Facility -- CSIRO,
P.O. Box 76, Epping NSW 1710, Australia}
\altaffiltext{4}{Columbia Astrophysics Laboratory, Columbia University, 550 West 120th Street, New York, NY 10027, USA}
\altaffiltext{5}{Dept. of Physics and Astronomy, University of British
Columbia, 6224 Agricultural Road, Vancouver, BC V6T 1Z1 Canada}
\altaffiltext{6}{Universit\`a degli Studi di
Cagliari, Dipartimento di Fisica, SP Monserrato-Sestu km
0.7, 09042 Monserrato, Italy}
\altaffiltext{7}{NAIC, Arecibo Observatory, HC03 Box 53995, PR 00612, USA}
\altaffiltext{8}{National Centre for Astrophysics, P.O. Bag 3, Ganeshkhind, Pune 411007, India
}
\altaffiltext{9}{Massachusetts Institute of Technology, Haystack Observatory, Westford, MA 01886, USA}

\begin{abstract}

We have investigated the eclipse of the 23 ms pulsar,
PSR~J0737$-$3039A, by its 2.8 s companion PSR~J0737$-$3039B in the
recently discovered double pulsar system using data taken with the
Green Bank Telescope at 820 MHz. We find that the pulsed flux density
at eclipse is strongly modulated with the periodicity of the 2.8 s
pulsar. The eclipse occurs earlier and is deeper at those rotational
phases of B when its magnetic axis is aligned with the line of sight
than at phases when its magnetic axis is at right angles to the line
of sight. This is consistent with the eclipse of A being due to
synchrotron absorption by the shock-heated plasma surrounding B, the
asymmetry arising from the higher plasma densities expected in the B
magnetosphere's polar cusps.

\end{abstract}


\section{Introduction} \label{sec:intro}

The two pulsars in the recently discovered double pulsar binary
system, \psra and \psrb (hereafter simply ``A'' and ``B'') have
periods of $P_A = 23$~ms and $P_B = 2.8$~s. They are in a 2.4-hr
mildly-eccentric orbit that we view nearly edge-on, with an
inclination angle of $87\deg \pm3\deg$ (Burgay et al.~2003, Lyne et
al.~2004). The phenomenology exhibited by this system is extremely
rich.  The flux density of the B pulsar varies dramatically and
systematically around the orbit (Lyne et al.~2004, Ramachandran et
al.~2004, McLaughlin et al.~2004), indicating significant interaction
between the two pulsars. While the flux density of the A pulsar is
constant for most of the orbit, for $\sim$ 30~s around its superior
conjunction, A is eclipsed by the magnetosphere of B (Lyne et
al.~2004).  This eclipse is asymmetric, with the flux density of A
decreasing more slowly on ingress than it increases on egress. The
duration of the eclipse is mildly dependent on radio frequency, with 
eclipses lasting longer at lower frequencies (Kaspi et al.~2004).

Theoretical models by Arons et al.~(2004) and Lyutikov (2004) explain
these eclipse properties in the context of synchrotron absorption of
A's radio emission by the shock-heated plasma surrounding and
containing B's magnetosphere. Because B's magnetosphere will respond
differently to the pressure of the relativistic wind of A at different
rotational phases of B, one prediction of these models is that the
density of absorbing plasma, and hence the properties of A's eclipse,
will depend on the rotational phase of B.  In their early analysis of
the exploratory observations performed at the 100-m Green Bank
Telescope (GBT) in 2003 December and 2004 January, Kaspi et al.~(2004)
found no evidence for modulation of the eclipse properties with B's
rotational phase. We have analyzed this publically available
dataset and, in this Letter, show that the eclipse properties of A are
in fact strongly dependent upon the rotational phase of B.

\section{Observations and Analysis} \label{sec:obsandresults}

The double pulsar system J0737--3039 was observed at the GBT in 2003
December and 2004 January using receivers centered at at 427, 820 and
1400 MHz.  The 5-hr 820-MHz observation discussed here was acquired
with the GBT Spectrometer SPIGOT card with a sampling time of
40.96~$\mu$s and 1024 synthesized frequency channels covering a 50-MHz
bandwidth.  For further details of the observations and data
acquisition system, see Ransom et al.~(2004) and references therein.

The data were dedispersed and folded using freely available software
tools (Lorimer 2001) assuming the nominal dispersion measure of
48.9~cm$^{-3}$~pc (Burgay et al.~2003) and using ephemerides for A and
B from Lyne et al.~(2004). We folded the data with 512 phase bins and,
for each A pulse, calculated the mean pulsed flux density in two
on-pulse phase windows centered on the two pulse components (Burgay et
al.~2003). From these, we subtracted the baseline level calculated
from two off-pulse windows. The on- and off-pulse regions contained 95
and 320 phase bins, respectively. For each orbit, we created a light
curve of A's pulsed flux density with time by averaging every 12
pulses, for an effective time resolution of $\sim$~0.27~s. Orbital
phases were calculated using the modified Julian dates of pulse
arrival at the Solar-System barycenter and the pulsar
ephemeris. Analysis of these light curves reveals that, for the
majority of the orbit, the pulsed flux density of A is constant within
measurement uncertainties, exhibiting no obvious orbital
phase-dependent variations. However, close to superior conjunction
(i.e. orbital phase 90$\deg$, where orbital phase is defined as the
longitude from the ascending node), the pulsed flux density of A
varies dramatically.
 
\medskip
\epsfxsize=9truecm
\epsfbox{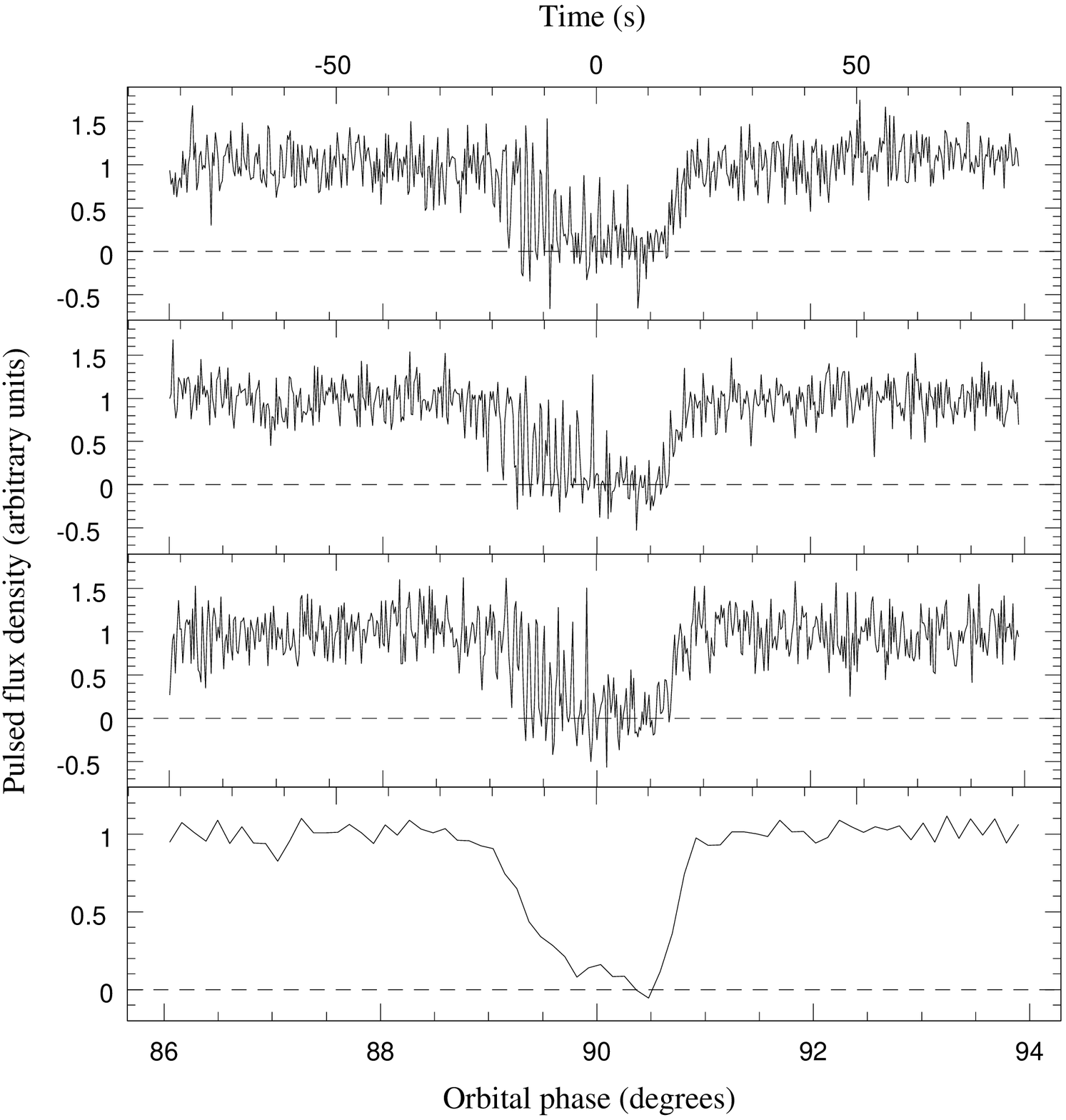}
\figcaption{ \label{fig:eclipse1}
 The pulsed flux density of A versus time (with respect to superior
conjunction) and orbital phase for (top three panels) the three
eclipses in the 820-MHz observation and (bottom panel) all three
eclipses summed. In the individual eclipse light curves, every 12
pulses have been averaged for an effective time resolution of
$\sim$~0.27~s. Every
100 pulses have been averaged to create the lower, composite light
curve for an effective time resolution of $\sim$~2.3~s. Pulsed flux
densities have been normalized such that the pre-eclipse average flux
density is unity.  }
\bigskip
  
Figure~\ref{fig:eclipse1} shows the light curves of A for all three
eclipses included in the 820-MHz observation. In the lower panel, we
also show a composite light curve, averaged over all three eclipses
and over 100 A pulses, for an effective time resolution of
$\sim$~2.3~s. Our measurements of the eclipse duration and of ingress
and egress shapes from this composite light curve are consistent with
the results of Kaspi et al.~(2004), who averaged data in 2 s
intervals. They calculated an eclipse duration (defined by the full
width at half maximum of the light curve) of 27~s, which was used to place
a limit of 18,600~km on the size of the eclipsing region.  They also
found that eclipse ingress takes roughly four times longer than egress
and that the eclipse duration is frequency dependent, lasting longer at
lower frequencies. 

\medskip
\epsfxsize=9truecm
\epsfbox{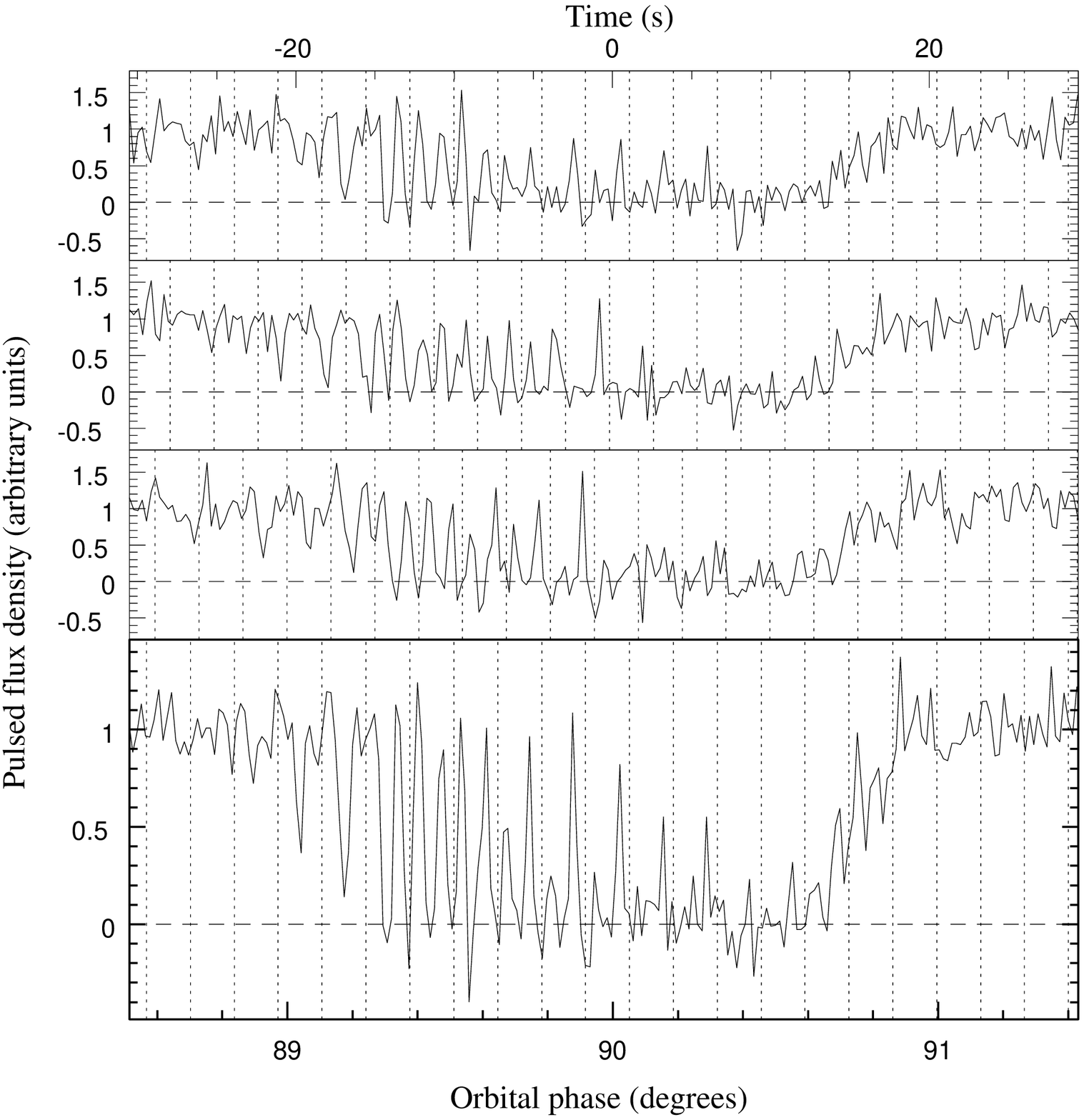}
\figcaption{
\label{fig:eclipse2}
As in Figure~\ref{fig:eclipse1}, the top three panels show the light
curves for the three individual eclipses, but covering a narrower
range of orbital phase. The vertical dashed lines indicate the measured arrival
times of the pulses of B.  The lower panel shows the light curve
averaged over all three eclipses of A, with the second and third
eclipses shifted by $<\pm P_B/2$ so that the B pulses arrive at the
same phase as during the first eclipse. Because limited
signal-to-noise ratio prohibits us from summing over a smaller number of
pulses, it is likely that the true structure of the flux density peaks
and dips is finer than it appears in this figure.
}
\bigskip

Inspection of the top three panels of Figure~\ref{fig:eclipse1} shows
that the rise and fall of the flux density during eclipse is not
monotonic. In fact, up until orbital phases approaching 90$\deg$, the
light curves of A show amplitude peaks consistent with pre-eclipse
levels.  Figure~\ref{fig:eclipse2} shows an expanded view of
Figure~\ref{fig:eclipse1} covering a smaller range of orbital phases
centered on the eclipse. On this plot, we also indicate the measured
barycentric arrival times of the pulses of the 2.8~s pulsar B. This
demonstrates clearly that the pulsed flux density of A is modulated in
synchronism with $0.5~P_B$.  In all three eclipses, we see negative dips
in pulsed flux density occurring first at a time about 0.5~$P_B$ out
of phase with the B pulses, and later also at a time in phase with the
B pulses.  In order to determine more sensitively how the flux density
of A depends upon the rotational phase of B, we shifted the light
curves of the second and third eclipses by up to $\pm~0.5~P_B$ so that
the phases of the B pulses were identical to those of the first
eclipse. We then summed the light curves of all three eclipses to
create the bottom panel of Figure~\ref{fig:eclipse2}.  Dividing each
2.8 s window of B's rotational phase into four equal regions, we can
calculate an average light curves for each region, as shown in
Figure~\ref{fig:eclipse3}. The light curves for each of the B pulse
phase windows vary smoothly.

Eclipse durations range from 20 s to 34 s for the four B phase
windows, calculated, as in Kaspi et al.~(2004), to be the full width
at half maximum of the eclipse.  Ingress occurs first at B phase 0.5
and then at 0.0, when, assuming B is an orthogonal rotator, the
magnetic axis of B is aligned with the line of sight to A.  It occurs
later and more gradually for the other two phase windows centered on
phases 0.25 and 0.75, when the magnetic axis of B is at right angles
to the line of sight.  The orbital phases of egress are similar for
all B pulse phases.  For light curves at B phases centered on 0.0 and
0.5, the eclipses are essentially symmetric, with minimum beginning at
orbital phases near 89.4$\deg$ and ending at orbital phases near
90.6$\deg$. The light curves for B phases centered on 0.25 and 0.75
are less symmetric, with the flux density minimum occurring briefly,
around orbital phases $90.2\deg - 90.5\deg$.  Only at around orbital phase
90.5$\deg$ is flux density of all phases of B consistent with being
zero.

It is
important to recognize that the features we see in the 820-MHz data
cannot be due to confusion with the single pulses of B
(e.g. McLaughlin et al.~2004) since the pulsed features seen in Fig.~2
are not in phase with the radio pulses of B, and, in addition, the latter,
which have width of $\sim 50$~ms, would
be removed by the on-pulse/off-pulse subtraction procedure used
to determine A's flux density.

\medskip
\epsfxsize=9truecm
\epsfbox{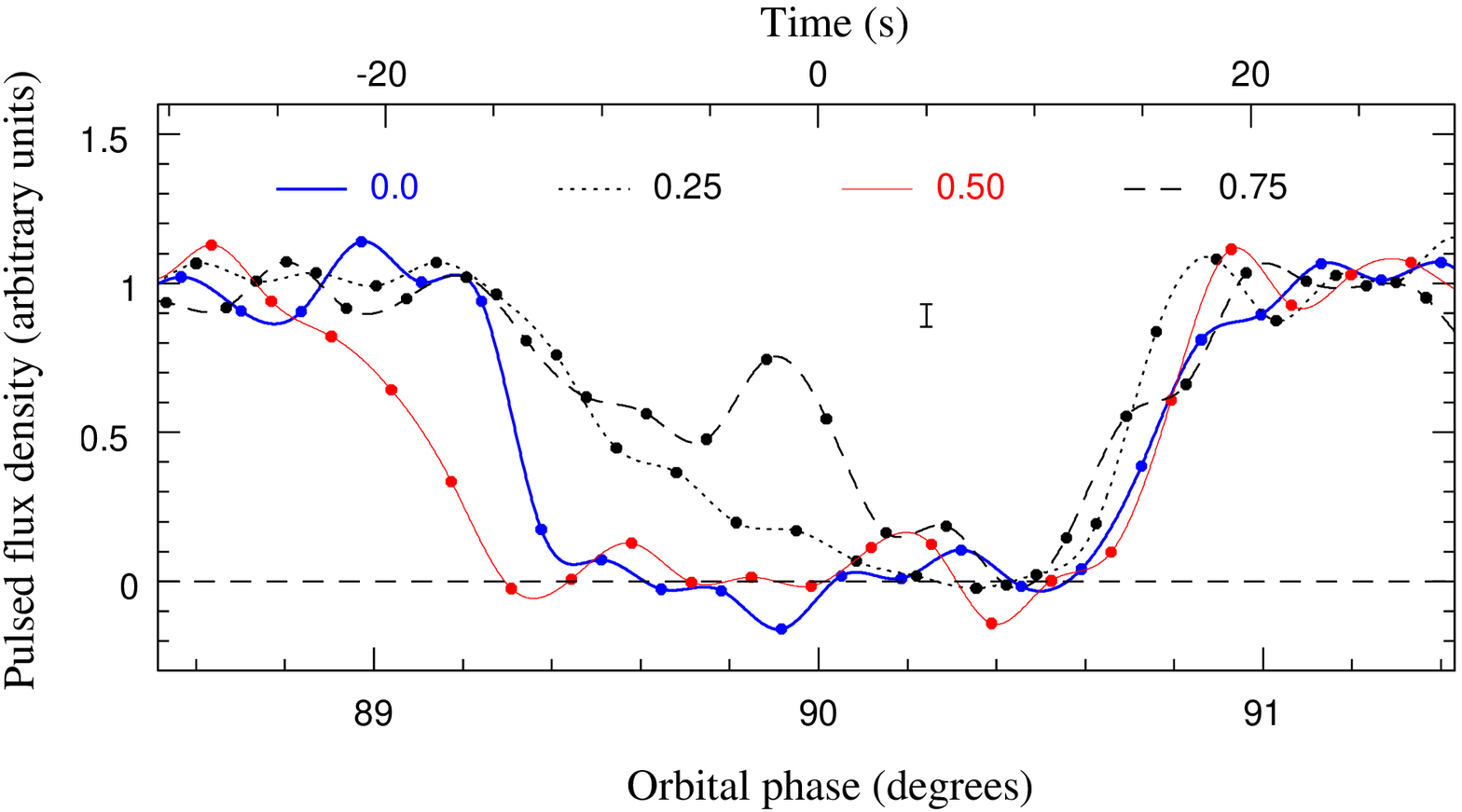}
\figcaption{
\label{fig:eclipse3}
Averaged light curves for four regions of B pulse phase, with a smooth
curve calculated with a spline fitting procedure drawn through the
individual flux density measurements. The four regions are centered on
pulse phases 0.0, 0.25, 0.50 and 0.75, with, for example, the curve
for 0.0 covering the 0.25 of pulse phase centered on B's radio
pulse. Figure~\ref{fig:eclipse4} shows the geometry of the system at
this B pulse phase. The vertical bar indicates the typical $\pm$ one-sigma
measurement error in each flux density.}
\bigskip

\section{Discussion} \label{sec:discussion}

Our analysis of these data differs from that of Kaspi et al.~(2004) in
one significant respect.  While Kaspi et al.~(2004) calculated light
curves with 2 s time resolution, we have integrated over only 12 A
pulses, affording us $\sim$ 7 times better time resolution. Kaspi et al. (2004)
did detect some emission from A on short time scales during eclipse, but
their 2 s integration time was not sufficient to detect the B modulation.
We have shown that the eclipse duration
varies considerably with
B phase and, hence, that the properties of the occulting medium are in
fact different in different regions of B's magnetosphere.

The observed eclipse phenomenology can be understood in the context of
a model invoking synchrotron absorption of the radio emission from A
in a magnetosheath surrounding B's rotating magnetosphere (Arons et
al.~2004, Lyutikov 2004).  Much of the rotational energy of the
faster-spinning A pulsar is emitted in the form of a relativistic,
magnetized wind composed primarily of electron-positron pairs.  The
rotational energy loss rate of A ($6\times10^{33}$~erg~s$^{-1}$) is
$\sim$ 3600 times greater than that of the slower B pulsar. Because the
dynamic pressure of A's wind is balanced against B's magnetic pressure
only at a point well inside the light cylinder of B, A's wind confines
the B magnetosphere on the side facing A, with a long, magnetospheric
tail extending behind B. The collision of A's wind with the
magnetosphere of B leads to the formation of a bow shock consisting of
hot, magnetized plasma. This plasma will surround the cometary shaped
magnetosphere of B, creating what Arons et al.~(2004) call a
``magnetosheath''. In Figure~\ref{fig:eclipse4}, we show a schematic
of the interaction of A's relativistic wind with B's magnetosphere.
The rotation of B inside the magnetosheath is expected to modulate the
shape of the sheath and to produce cusps within it,
leading to higher plasma densities in the cusps. Note that the
plasma will not reach all the way to the polar caps of B, but is confined
at the ``magnetopause'', where
the plasma pressure equals the magnetic pressure.
An excellent description
of the shape of the magnetosphere and of the polar cusps, is given
in Arons \& Lea (1976). 

\medskip
\epsfxsize=9truecm
\epsfbox{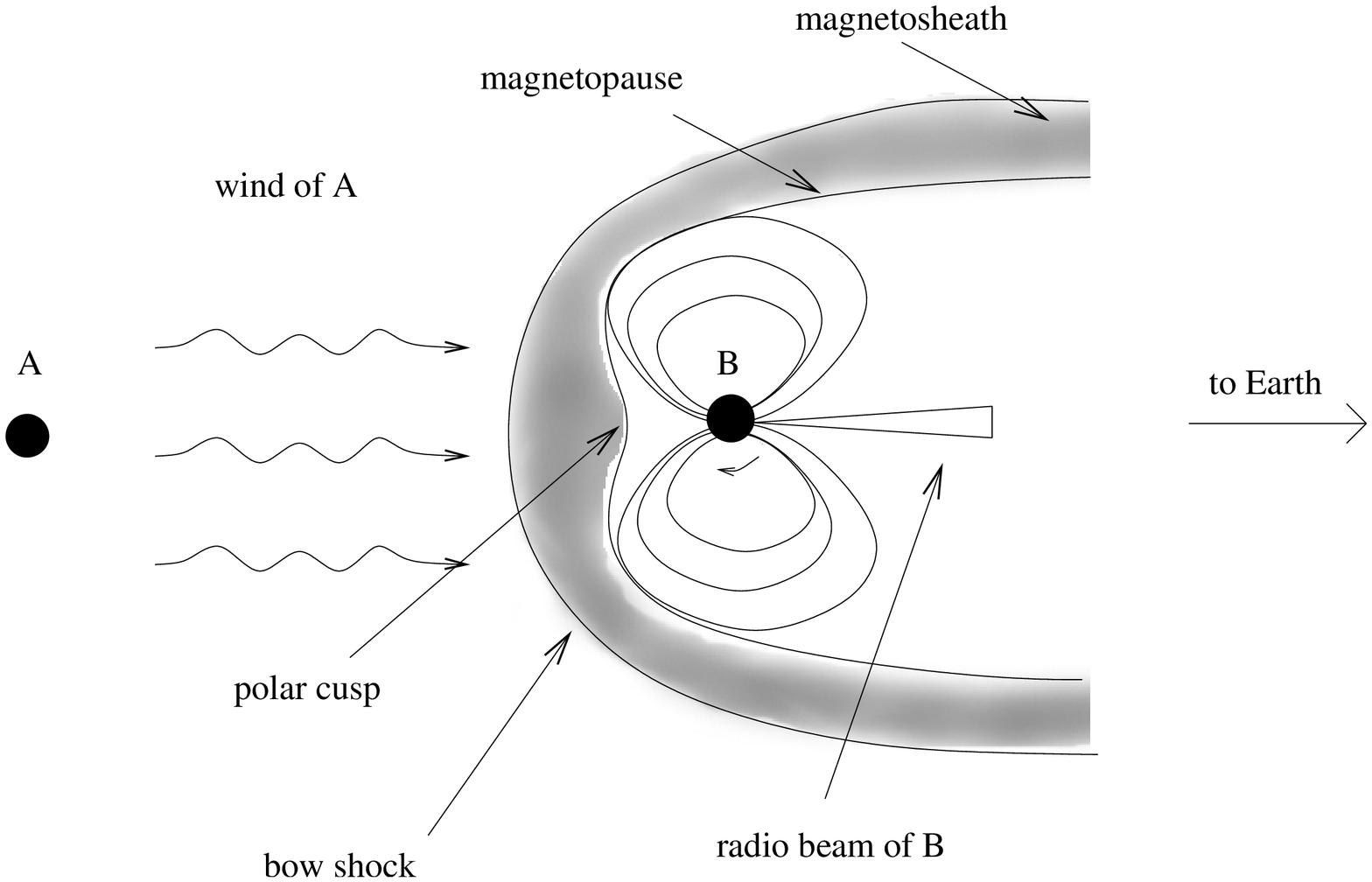}
\figcaption{
\label{fig:eclipse4}
Cartoon (not to scale) showing the interaction between the
relativistic wind of A and the magnetosphere of B
when the radio beam of B is pointing towards the Earth (i.e. pulse phase 0.0).
For a
more detailed picture of this interaction, see Figure~2 of Arons et
al.~(2004).}
\bigskip

As already discussed qualitatively by Kaspi et al.~(2004) and
quantitatively by Arons et al.~(2004) and Lyutikov (2004), synchrotron
absorption in such a magnetosheath can explain many of the properties
of the A eclipse, including the sharp edge and the mild frequency dependence.
If B is rotating prograde to the orbital velocity, it can also
explain the asymmetry between eclipse ingress and egress. The observed
changes in eclipse properties with B pulse phase give further support
to the model. The pulses from A are first absorbed at the two B pulse
phases (i.e. 0.0 and 0.5) at which the magnetic axis of B is aligned
with the line of sight to A. This can be attributed to the higher
density of absorbing plasma in B's polar cusps. Note that the pulses
at B pulse phase 0.50 (i.e. when the radio-bright pole is pointed
towards A) are absorbed before those at B pulse phase 0.0.  When the
line of sight to A is at a right angle to B's magnetic axis, the
pulses of A travel through regions of higher plasma density later in
the orbit and start fading closer to superior conjunction, where the
opacity of the magnetosheath tail could also become predominant.  As
shown in Figure~\ref{fig:eclipse3}, the ingress phase and eclipse
duration for these two B phases (i.e. 0.25 and 0.75) differ, with
absorption happening later at phase 0.75.  This may be due to the
rotationally-induced asymmetry predicted by Arons et al.~(2004).

The above discussion, and the simulations of Arons et al.~(2004),
assume that B is an orthogonal rotator.  As Demorest et al.~(2004) and
Arons et al.~(2004) discuss, it is quite likely that this is the case
as the wind torque from A should have aligned B's rotation axis with
the orbital angular momentum.  Furthermore, it is very difficult to
explain the similarity between the light curves at B pulse phases 0.0
and 0.5 and phases 0.25 and 0.75 if B is not nearly orthogonal.

From the measured eclipse durations, it is clear that the size of the
eclipsing region is much smaller than the light cylinder of
B. However, because the plasma density varies greatly depending on the
rotational phase of B, determining the extent and density of the
eclipsing region is not trivial and will require detailed modeling of
the plasma density variations within B's magnetosphere.  Due to the
effects of relativistic periastron advance and geodetic precession,
the durations and morphologies of the eclipse and their dependence on
B rotational phase should change significantly with time.  Forthcoming
observations will allow us to sensitively probe the dynamics of the
wind/magnetosphere interaction and of the geometry of the system.
These observations, at multiple frequencies, will also allow us to
test the hypothesis of Arons et al.~(2004) that there should be no
eclipse at frequencies above 5~GHz, already supported by the measurements of
Kaspi~et~al.~(2004).

Because pulsars emit a large portion of their spin-down energy in the
form of a relativistic wind, understanding these winds is crucial for
forming a complete picture of the pulsar energy budget. However,
because the winds cannot be directly observed, progress in this field
has been limited to studies of pulsar wind nebulae and bow shocks.  If
the eclipses of A are indeed caused by synchrotron absorption in B's
magnetosheath, both Arons et al. (2004) and Lyutikov (2004) reach the
very interesting conclusion that the density of A's wind must be at
least four orders of magnitude greater than is expected given
currently accepted models of pair creation.  The Arons et al.~(2004)
model also implies a very low wind magnetization.  However, in our
paper describing the drifting features observed in the single pulses
of B (McLaughlin et al.~2004), we concluded that that most of the
spin-down energy is carried by the Poynting flux rather than by
energetic particles. Clearly, self-consistent models which can explain
both of these phenomena are needed. The rich phenomenology offered by
the double pulsar system will offer us new ways to explore the physics
of these winds on much smaller scales than has been possible with
other systems. Eclipse measurements of the double pulsar system may
also be important for constraining pulsar pair creation scenarios.

\acknowledgments

We thank A. Spitkovsky for a very useful discussion which motivated us
to look for this effect and D. Melrose, F. Graham Smith and J. Arons for helpful
discussions.
We thank the National Radio Astronomy Observatory for
making these observations publically available.  The National Radio
Astronomy Observatory is facility of the National Science Foundation
operated under cooperative agreement by Associated Universities,
Inc. IHS holds an NSERC UFA and is supported by a Discovery Grant. DRL
is a University Research Fellow funded by the Royal Society.  NDA, AP
and MB received support from the Italian Ministry of University and
Research (MIUR) under the national program {\it Cofin 2003}.

{}

\end{document}